\documentclass[aps,prb,showpacs,twocolumn,10pt,longbibliography]{revtex4-1}

\usepackage{graphicx}
\usepackage{amssymb}
\usepackage{amsmath}
\usepackage{color}
\usepackage{dsfont}
\usepackage{paralist}
\usepackage{bm}
\usepackage{verbatim}

\usepackage{amsfonts}
\usepackage{graphicx}
\usepackage{bm}
\usepackage{color}

\def\cH{\hat{\cal H}}

\def\bJ{{\bf J}}
\def\bn{{\bf n}}
\def\bk{{\bf k}}

\def\bq{{\bf q}}
\def\bp{{\bf p}}
\def\br{{\bf r}}

\def\ve{\varepsilon}

\def\hbsigma{\hat{\boldsymbol \sigma}}

\def\hpsi{\hat \psi}

\def\hsigma{\hat\sigma}

\DeclareMathOperator{\arcsinh}{arcsinh}

\begin{document}

%%%%%%%%%%%%%%%%%%%%%TITLE%%%%%%%%%%%%%%%%%%%%%%%%%%%%%%%%%%%%%%%%%%%%%%%%%%%%%%%%%%%
\title{Conductivity of a Weyl semimetal with donor and acceptor impurities}

\author{
Ya.I.~Rodionov$^1$ and S.V.~Syzranov$^{2,3}$
}
\affiliation{
$^1$Institute for Theoretical and Applied Electrodynamics RAS, 125412 Moscow, Russia \\
$^2$Physics Department, University of Colorado, Boulder, Colorado 80309, USA\\
$^3$Center for Theory of Quantum Matter, University of Colorado, Boulder, Colorado 80309, USA
}

%%%%%%%%%%%%%%%%%%%%%%%%%%ABSTRACT%%%%%%%%%%%%%%%%%%%%%%%%%%%%%%%%%%%%%%%%%%%%%%%%%%%%%%%%%%%%%%%%%%%%%%%%%%%%
\begin{abstract}
  We study transport in a Weyl semimetal with donor and acceptor impurities.
  At sufficiently high temperatures transport is dominated by electron-electron interactions,
  while the low-temperature resistivity comes from the scattering of quasiparticles on screened impurities.
  Using the diagrammatic technique, we calculate the conductivity $\sigma(T,\omega,n_A,n_D)$ in the
  impurities-dominated regime as a function of temperature $T$, frequency $\omega$, and the concentrations
  $n_A$ and $n_D$ of donors and acceptors and discuss the crossover behaviour between the regimes
  of low and high temperatures and impurity concentrations.
  In a sufficiently compensated material [$|n_A-n_D|\ll(n_A+n_D)$] with a small effective fine
  structure constant $\alpha$,
  $\sigma(\omega,T)\propto T^2/(T^{-2}-i\omega\cdot\text{const})$ in a wide interval of temperatures.
  For very low temperatures or in the case of an uncompensated material
  the transport is effectively metallic. We discuss experimental conditions necessary for realising each regime.
\end{abstract}
%%%%%%%%%%%%%%%%%%%%%%%%%%%%%%%%%%%%%%%%%%%%%%%%%%%%%%%%%%%%%%%%%%%%%%%%%%%%%%%%%%%%%%%%%%%%%%%%%%%%%%%%%%%%%%%%


%%%%%%%%%%%%%%%%%%%%%%%%%%%%%%%%%%%%%%%%%%%%%%%%%%%%%%%%%%
%71.55.Ak	Metals, semimetals, and alloys
%72.10.-d	Theory of electronic transport; scattering mechanisms
%78.40.Kc	Metals, semimetals, and alloys
%72.15.Lh	Relaxation times and mean free paths (in metals and alloys)
%72.80.Ng	Disordered solids (under transport in specific materials)
%72.80.Vp	Electronic transport in graphene

\date{\today}

\maketitle

%%%%%%%%%%%%%%%%%%%%%%%%%%%%%%%%%%%%%%%%%%%%%%%%%%%%%%%%%%%%%%%%%%%%%%%%%%%%%%%%%%%%%

\section{Introduction}

Weyl\cite{Burkov:WeylProp,Wan:WeylProp,Hasan:ARPESobs,TaAsFirst} and
Dirac\cite{Liu:Na3Bi,Hasan:Cd3As2,Cava:Cd3As2,Yazdani:Cd3As2,Liu:Cd3As2} semimetals,
3D materials with Weyl and Dirac quasiparticle dispersions,
are expected to display a plethora of unconventional previously unobserved transport phenomena
such as the absence of localisation by smooth non-magnetic disorder\cite{Wan:WeylProp,RyuLudwig:classification}
or disorder-driven phase
transitions\cite{Fradkin1,Fradkin2,GoswamiChakravarti,Syzranov:Weyl,Brouwer:WSMcond,Moon:RG,Herbut}
similar to the localisation transition in high-dimensional semiconductors\cite{Syzranov:unconv}.

%First WSMs have been identified quite recently in ARPES experiments\cite{Hasan:ARPESobs,TaAsFirst},
%as well as closely related Dirac semimetals (DSMs),
%3D materials with four-band Dirac quasiparticle dispersion, that can be considered as a partial case of WSMs.

The character of
transport phenomena, observable in such systems, dramatically depends on the nature and amount
of quenched disorder.
For instance, short-range disorder has been predicted to strongly renormalise the properties
of long-wave quasiparticles\cite{DotsenkoDotsenko,Fradkin1,Fradkin2,LudwigFisher,Nersesyan:dwave,GoswamiChakravarti,AleinerEfetov,Syzranov:Weyl,OstrovskyGornyMirlin,Moon:RG,Herbut,RoyDasSarma},
leading to a disorder-driven phase transition, that is expected to manifests itself, e.g.,
in a critical behaviour of the conductivity\cite{Fradkin1,Syzranov:Weyl,Brouwer:WSMcond}
or the density of states\cite{Herbut,Syzranov:unconv}
near a critical disorder strength. However, such transition does not exist
for Coulomb impurities, which are more likely to dominate
transport in such systems (while the critical behaviour in the density of states is still observable\cite{Syzranov:unconv}).

Transport in Weyl semimetals (WSMs) with charged scatterers has been extensively addressed in the literature
in the limits of sufficiently low and high doping
levels and temperatures\cite{Skinner:WeylImp,BurkovHookBalents,Ominato:Coulomb,HwangSarma,Lundgren:thermoelectric,Adam:magnetoresistance}.
Coulomb impurities have been predicted to manifests themselves, e.g., in the temperature dependency\cite{HwangSarma}
$\sigma\sim T^4$ of conductivity at high temperatures.
For sufficiently low temperatures and
levels of doping, fluctuations in the concentration of charged impurities
lead to the formation of electron and hole puddles, that
determine the minimal conductivity of a WSM\cite{Skinner:WeylImp}.
For a sufficiently small amount of disorder, it is expected that resistivity is dominated
by electron-electron interactions\cite{Abrikosov:metals,CompleteRubbish,BurkovHookBalents},
that lead to a finite resistivity even in disorder-free samples.

%%%%%%%%%%%%%%%%%%%CASES%%%%%%%%%%%%%%%%%%%%%%%%%%%%%%%%%%%%%%%%%%%%%%%%%%%%%%%%%%%%%%%%%%%%%%%%%%%%%%%%%%
%Transport in WSMs and DSMs has been extensively studied in the literature and has been addressed theoretically
%in a number of limiting cases: in disorder-free samples\cite{CompleteRubbish,BurkovHookBalents},
%for short-range
%disorder\cite{Fradkin1,CompleteRubbish,BurkovHookBalents,Ominato:WeylDrude,Syzranov:Weyl,Brouwer:WSMcond,AltlandBagets,Lu:corrections,Ziegler}  at zero and finite frequencies (cf. also Ref.~\onlinecite{BiswasRyu}), and
%in the dc limit for Coulomb scatterers for sufficiently low and high doping
%levels and temperatures\cite{Skinner:WeylImp,BurkovHookBalents,Ominato:Coulomb,HwangSarma,Lundgren:thermoelectric}.
%If the doping level is very low, the fluctuations of the impurity concentration
%lead to the formation of electron and holes puddles, that determine the minimal conductivity of a WSM,
%as predicted in Ref.~\onlinecite{Skinner:WeylImp}.
%%%%%%%%%%%%%%%%%%%%%%%%%%%%%%%%%%%%%%%%%%%%%%%%%%%%%%%%%%%%%%%%%%%%%%%%%%%%%%%%%%%%%%%%%%%%%%%%%%%%%%%%%%%

\begin{figure}[b]
	\centering
	\quad\quad\includegraphics[width=0.43\textwidth]{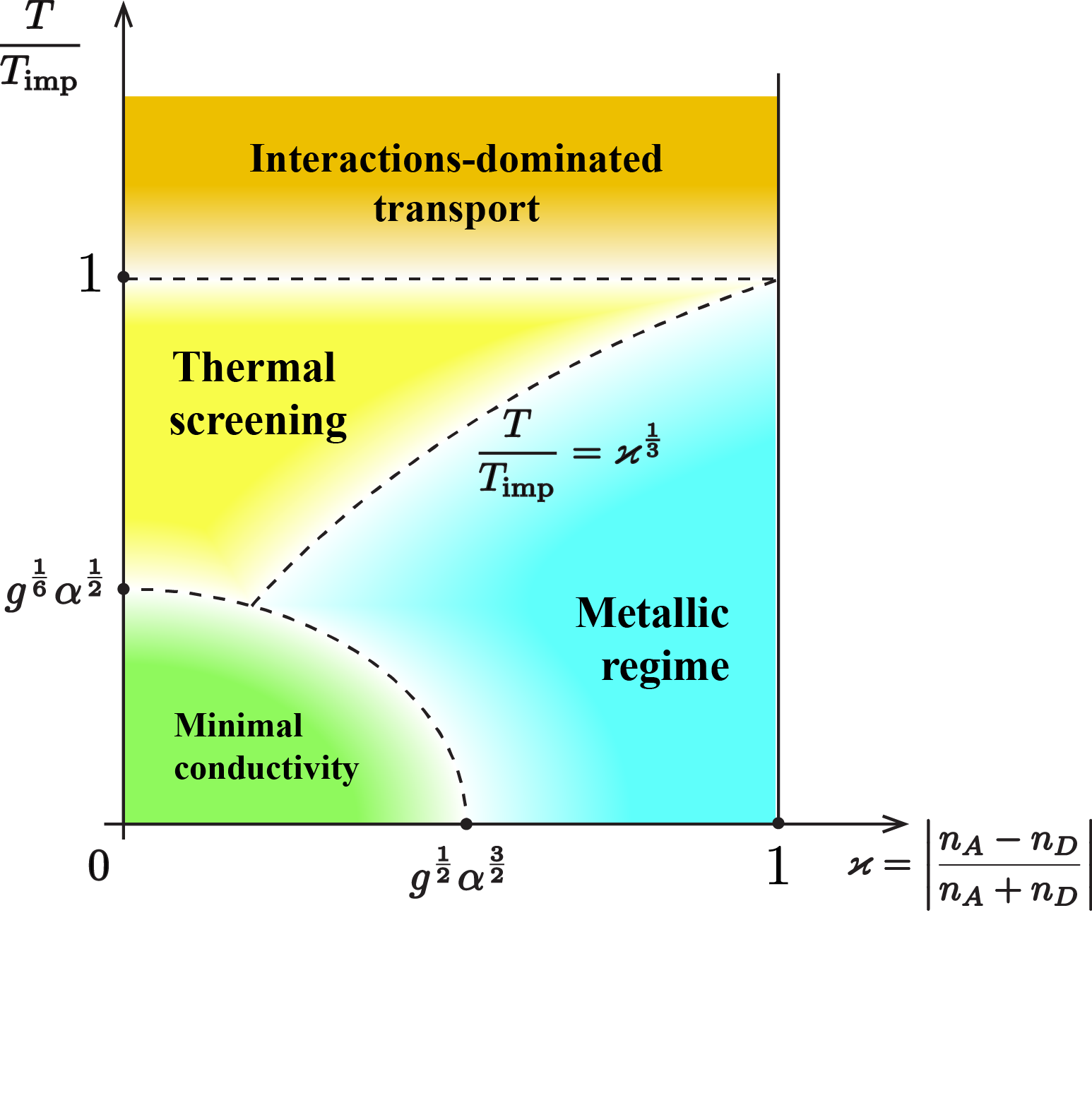}
	\caption{
	\label{Fig:Regimes}
	(Colour online) The temperature $T$ \{in units $T_{\text{imp}}=[(n_A+n_D)v^3/g]^{1/3}$\}
	vs. the compensation parameter $\varkappa=\left|{n_A-n_D}\right|/({n_A+n_D})$	
	diagram for a Weyl semimetal with donor and acceptor impurities, where
	$n_A$ and $n_D$ are the concentrations of acceptor and donor impurities
	and $g$-- the degeneracy of the Weyl point.
	}
\end{figure}

Currently it still remains to be investigated which of these phenomena and regimes of conduction
can be realised in WSMs,
under what conditions, and which of them display transport features specific to Weyl
materials.
Indeed, %in a disorder-free WSM the chemical potential is located exactly at the Weyl point.
charged impurities, intrinsically present in realistic materials,
lead to a finite chemical potential $\mu$ (measured
from the Weyl point), making WSM similar to a usual metal in terms of transport
properties at low temperatures $T\ll|\mu|$.
Signatures of Weyl (Dirac) quasiparticles scattered by
Coulomb impurities are expected\cite{HwangSarma} to be detectable at higher temperatures, $T\gg|\mu|$.
However, the rate of quasiparticle scattering due to electron-electron interactions also grows
with temperature\cite{Abrikosov:metals,CompleteRubbish} and may prevail over the scattering on impurities.

Another question, that deserves investigation, is the dependency of the ac conductivity $\sigma(\omega)$
of a WSM on frequency $\omega$, as it can be used to directly measure
the quasiparticle scattering time $\tau$ as a function of temperature or doping
in the frequency range $\omega\sim\tau^{-1}$
and thus can provide information on the mechanisms of transport and the nature of disorder in
a material.

%\subsection*{Summary of the results}

In this paper we study the conductivity $\sigma(T,\omega,n_A,n_D)$ of a WSMs with donor (positively charged)
and acceptor (negatively charged) impurities as a function of temperature $T$, frequency $\omega$, and the
concentrations $n_A$ and $n_D$ of donors and acceptors.

Fig.~\ref{Fig:Regimes} summarises different regimes
of transport that can be achieved in a WSM by varying temperature and the concentrations of donors
and acceptors.
The frequency and temperature
dependencies of conductivity in the ``metallic regime'' resemble those of a usual metal.
The ``interactions-dominated transport'' is dominated by interactions and weakly depends on disorder.
In the regime of ``thermal screening'' the conductivity and the
screening of impurities is determined by electrons thermally
excited from the valence to the conduction band, and the dc conductivity is strongly temperature-dependent,
as predicted for small $\varkappa$ in Ref.~\onlinecite{HwangSarma}.
At low $T$ and $\varkappa=\left|{n_A-n_D}\right|/({n_A+n_D})$
strong fluctuations of the disorder potential lead to the formation of electron
and hole puddles that determine the ``minimal conductivity'' introduced in Ref.~\onlinecite{Skinner:WeylImp}.

Focussing on the disorder-dominated transport away from strong random-potential fluctuations
(the ``metallic'' and ``thermally screening'' regions in Fig.~\ref{Fig:Regimes}),
we calculate the conductivity $\sigma(T,\omega,n_A,n_D)$ explicitly and discuss the crossover
behaviour between different regimes.

Our results apply both to Weyl semimetals and to Dirac semimetals, as the latter may be
considered as Weyl semimetals with merging pairs of Weyl points.

%The dependency of the conductivity on frequency $\omega$, to our knowledge unadressed
%Also, ac conductivity of a WSM with charged impurities has so far evaded researchers' attention,
%despite the dependency of the conductivity on the frequency may be used to probe the quasiparticle
%scattering time for frequencies of the order of the scattering rate.

The paper is organised as follows. In Sec.~\ref{Sec:model} we
introduce the model for a Weyl semimetal
with dopant impurities.
Sec.~\ref{Sec:densities} deals with relations between the concentration of dopants,
chemical potential, and the fluctuations of the electron density.
In Sec.~\ref{Sec:scattering} we discuss the
mechanisms of quasiparticle scattering and discuss the conditions under which the resistivity
is disorder- and interactions-dominated. We evaluate the conductivity
explicitly in Sec.~\ref{Sec:GenExpr} and discuss the crossovers between various regimes of transport
in Secs.~\ref{Subsec:LowT}-\ref{Subsec:dcLim}.
In Sec.~\ref{Sec:conclusion} we summarise our results and discuss the experimental conditions
necessary for realising each regime.

%%%%%%%%%%%%%%%%%%%%%%%%%%%%%%%%%%%%%%%%%%%%%%%%%%%%%%%%%%%%%%%%%%%%%%%%%%%%%%%%%%%%%%%%%%%%
%%%%%%%%%%%%%%%%%%%%%Table Summary of previous results%%%%%%%%%%%%%%%%%%%%%%%%%%%%%%%%%%%%%%
%\begin{table}
%\begin{tabular}{p{2cm} || p{3cm}| p{3cm}}
%	  & $T \gg \mu$ & $T\ll \mu$ \\ \hline\hline
%	%%%%%%%%%%%%%%%FIRST ROW%%%%%%%%%%%%%%%%%%%%%%%
%	Short-range-correlated
%    disorder	& $\sigma= const$ \par
%	(Refs.~\onlinecite{CompleteRubbish,Ominato:WeylDrude,Syzranov:Weyl,Burkov:WeylProp})
%	& $\sigma=const$ \par
%	(Refs.~\onlinecite{CompleteRubbish,Ominato:WeylDrude,Syzranov:Weyl,Burkov:WeylProp})
%	\\ \hline
%    %%%%%%%%%%%%%%SECOND ROW%%%%%%%%%%%%%%%%%%%%%%%
%	Coulomb impurities & $\sigma\propto T^4$ \par (This paper)
%	& $\sigma\propto \mu$ \par
%	(Ref.~\onlinecite{BurkovHookBalents})
%\end{tabular}
%\caption{Conductivity
%of a weakly-disordered Weyl semimetal
%as a function of temperature and chemical potential
%for different temperatures
%and disorder models. The temperature or the chemical potential are assumed sufficiently
%large to neglect the rare-regions effects\cite{Nandkishore:rare}
%or electron and hole puddles created by the fluctuations
%of the impurity concentration\cite{Skinner:WeylImp}.}
%\end{table}
%%%%%%%%%%%%%%%%%%%%%%%%%%%%%%%%%%%%%%%%%%%%%%%%%%%%%%%%%%%%%%%%%%%%%%%%%%%%%%%%%%%%%%%%%%%
%%%%%%%%%%%%%%%%%%%%%%%%%%%%%%%%%%%%%%%%%%%%%%%%%%%%%%%%%%%%%%%%%%%%%%%%%%%%%%%%%%%%%%%%%%%

\section{Model}
\label{Sec:model}

The Hamiltonian of long-wavelength quasiparticles in a
WSM with charged impurities reads
\begin{subequations}
\begin{align}
	\cH &=\cH_0+\cH_{ee}+\cH_{\text{imp}}, \\
	\cH_0 &=\int \hpsi^\dagger(\br)v(\hbsigma\cdot\hat\bk)\:\hpsi(\br) d^3\br, \\
	\cH_{ee} &=\int\hpsi^\dagger(\br)\hpsi^\dagger(\br^\prime)\frac{e^2}{\kappa|\br-\br^\prime|},
	\hpsi(\br^\prime)\hpsi(\br) \:d^3\br\, d^3\br^\prime, \\
	\cH_{\text{imp}} &=\sum_i\int \hpsi^\dagger(\br)\frac{Z_i e^2 }{\kappa|\br-\br_i|}\hpsi(\br)\: d^3\br,
\end{align}
\end{subequations}
where $\cH_0$ is the Hamiltonian of free non-interacting Weyl fermions; $\hpsi^\dagger(\br)$
and $\hpsi(\br)$ are the fermion creation and annihilation operators, $v(\hbsigma\cdot\hat\bk)$ is the
quasiparticle dispersion, with $\hat\bk=-i\bf\nabla_\br$ being quasiparticle momentum, and $\hbsigma$-- the pseudospin operator;
$\cH_{ee}$ is the Hamiltonian of electron-electron interactions, $\varkappa$ being the dielectric constant;
the operator $\cH_{\text{imp}}$ describes the interaction between electrons and charged impurities, located at
random coordinates $\br_i$; $Z_i e$ is the charge of the $i$-th impurity.
Throughout the paper we set $\hbar=1$.
We consider two types of impurities: {\it acceptors}, with $Z_i=1$, and {\it donors}, with $Z_i=-1$.

Throughout the paper we assume for simplicity that the energies of bound states on the donor impurities are sufficiently
high, and those on the acceptor impurities are sufficiently low, so that donors are always ionised
and each acceptor always hosts an electron.
In a realistic material, however, electron occupation numbers
on dopant impurities may depend on the temperature and chemical potential.
Our results can be easily generalised to this more realistic case;
$n_A$ and $n_D$ should be understood then as the concentrations of impurities
with charges $-e$ and $+e$ respectively, explicitly dependent on temperature and dopant concentrations.

%In a realistic semiconductor donor and acceptor impurities may be charged or not, depending on
%the temperature chemical potential, and the energies of the bound states on the impurities.
%However, in this paper we assume for simplicity that the donors are always ionised, and each acceptor
%always hosts in electrons. In a more general situation, ``donors'' and ``acceptors'' should be understood
%as impurities with charges $-e$ and $+e$.

Due to the fermion doubling theorem\cite{Nielsen:doubling}, Weyl quasiparticle dispersion
is expected near an even number of points (Weyl points) in the first Brillouin zone.
However, quasiparticle scattering between different Weyl points can be neglected due to the smoothness
of the random potential
created by the charged impurities under consideration
and the due to long-range character of electron-electron interactions.
For simplicity, we assume identical quasiparticles dispersions near all Weyl points and at the
end of the calculation multiply the contribution of one point to the conductivity by a factor of
$g$, that accounts for the number of Weyl and spin degeneracy.

The strength of electron-electron interactions is characterised by the ``fine structure constant''
\begin{equation}
	\alpha=\frac{e^2}{v\kappa},
\end{equation}
which is assumed to be small, $\alpha\ll1$,
in this paper (for instance, in\cite{Cd3As2dielectric,Liu:Cd3As2} $Cd_3As_2$ $\alpha\sim 0.05$).
Throughout the paper we assume also, that the degeneracy $g$ is not very large, so that the
condition
\begin{equation}
	g\alpha\ll 1
\end{equation}
is fulfilled.

%%%%%%%%%%%%%%%%%%%%%%%%%%%%%%%%%%%%%%%%%%%%%%%%%%%%%%%%
\section{Charge-carrier density and impurity screening}
\label{Sec:densities}

In an undoped WSM the chemical potential is located at the Weyl point.
Adding donor and acceptor impurities with different concentrations leads to a finite
chemical potential $\mu$ (measured from the Weyl point) and an excess density $n(\mu)$ of electrons,
as compared to the undoped sample.
Charge neutrality of the material requires that
\begin{align}
	n_D-n_A=n(\mu,T),
	\label{neutrality}
\end{align}	
where %in a WSM, a 3D material with quasiparticle spectrum $\pm v|\bp|$ with degeneracy $g$,
the excess electron density is given by
\begin{align}
	n(\mu,T) &=g\sum_\pm\int\frac{d^3\bp}{(2\pi)^3}
	\left[f_0(\pm v|\bp|-\mu)-f_0(\pm v|\bp|)\right]
	\nonumber\\
	 &=g\frac{\mu^3+\pi^2\mu T^2}{6\pi^2 v^3},
	 \label{nofmu}
\end{align}
with $f_0(\varepsilon)=(e^{\varepsilon/T}+1)^{-1}$ being the Fermi distribution function.

{\it Fluctuations of electron density at low doping.}
Because impurities are located randomly, for very low densities $n(\mu,T)$ (i.e. for low $T$ and $\mu$)
the distribution of electron charge displays strong relative spatial fluctuations and cannot be assumed uniform.

Indeed, as discussed in Refs.~\onlinecite{Skinner:WeylImp,Shklovskii:book}, at small doping levels
the fluctuations of the concentration of
randomly located charged impurities lead to the formation of electron and hole
puddles.
In a WSM the characteristic depth of such puddles
is given by the energy scale\cite{Skinner:WeylImp}
\begin{equation}
	\Gamma\sim g^{-\frac{1}{6}}\alpha^\frac{1}{2}n_{\text{imp}}^\frac{1}{3}v,
	\label{Gamma}
\end{equation}
where $n_{\text{imp}}\equiv n_A+n_D$ is the total concentration of the (donor and acceptor) impurities.

Unless the concentrations of donors and acceptors are nearly equal, the characteristic energy of
the quasiparticles contributing to the conductivity can be estimated as
$\varepsilon\sim\max(T,|\mu|)\gtrsim v(n_{\text{imp}}/g)^\frac{1}{3}$ [using Eqs.~\eqref{neutrality}, \eqref{nofmu},
and $n_D-n_A\sim n_{\text{imp}}$] and significantly exceeds the scale
$\Gamma$, Eq.~\eqref{Gamma}.
Thus, electron and hole puddles (studied in Ref.~\onlinecite{Skinner:WeylImp})
can affect transport
only in {\it highly compensated} materials,
with nearly equal concentrations of donors and acceptors, $n_A\approx n_D$.

Using Eqs.~\eqref{neutrality}, \eqref{nofmu}, and \eqref{Gamma}, we find that electron and hole
puddles emerge only if
\begin{equation}
	\left|\frac{n_D-n_A}{n_D+n_A}\right|\ll g^\frac{1}{2}\alpha^\frac{3}{2},
	\label{CompensationHard}
\end{equation}
at low temperatures. Otherwise, relative fluctuations of electron distribution
may be considered small.

The condition
\begin{equation}
	\Gamma\lesssim \max(|\mu|,T)
\end{equation}
determines the boundary of the ``minimal conductivity'' regime (studied in Ref.~\onlinecite{Skinner:WeylImp})
in Fig.~\ref{Fig:Regimes}.
Because of the large depth
of electron and hole puddles in this regime, the conductivity is of the same order of magnitude
in the entire ``minimal conductivity'' region. Outside of it, the fluctuations of the electron density
are small, and the conductivity grows with temperature and chemical potential, as we find below.

\subsection*{Screening of impurities}

Finite chemical potential $\mu$ or temperature $T$ lead to the screening
of the charged impurities.
Assuming the distribution of electrons is sufficiently homogeneous [i.e. the condition \eqref{CompensationHard}
does not hold],
the screening radius $\lambda$ of the effective impurity potential
\begin{equation}
	\phi(\br)=\frac{e^2}{\varkappa r}e^{-\frac{r}{\lambda}}
	\label{Screened}
\end{equation}
is given in the Thomas-Fermi approximation
by\cite{Abrikosov:metals}
\begin{equation}
	\lambda^{-2}=4\pi e^2 n_\mu^\prime(\mu,T),
	\label{lambdaInit}
\end{equation}
with the excess density $n(\mu,T)$ of electrons given by Eq.~(\ref{nofmu}).

Using Eqs.~\eqref{lambdaInit} and \eqref{nofmu}, we obtain
\begin{equation}
	\lambda^{-2}=\frac{2g\alpha}{\pi v^2}
	\left(\mu^2+\frac{\pi^2}{3}T^2\right).
	\label{lambda}
\end{equation}
The Thomas-Fermi approximation is justified for small values of the fine structure constant, $\alpha\ll1$,
assumed in this paper. This condition ensures that the screening radius (\ref{lambda}) significantly exceeds the characteristic quasiparticle wavelength
$\lambdabar=v/\max(T,|\mu|)$. Also, under this condition one can neglect the renormalisation\cite{Syzranov:unconv}
of the low-energy quasiparticle properties from large momenta $k\gg\max(T,|\mu|)/v$.

{\it Thermal screening vs. Fermi sea.} Eq.~(\ref{lambda}) shows that
at sufficiently high temperatures, $T\gg|\mu|$,
the screening of the impurities is determined by electrons thermally excited from the valence
to the conduction band and is independent of the concentrations of the impurities.
In the opposite limit, $|\mu|\ll T$, the distribution of electron charge around an impurity
is equivalent to that in a usual metal\cite{Abrikosov:metals}. Thus, the boundary between
the ``thermal screening'' and ``metallic'' regimes in Fig.~\ref{Fig:Regimes} is given by the
condition $\mu\sim T$, which, according to Eqs.~\ref{neutrality} and \ref{nofmu}, is equivalent to
\begin{equation}
	T\sim v\left|\frac{n_D-n_A}{g}\right|^\frac{1}{3}.
\end{equation}

%%%%%%%%%%%%%%%%%%%%%SCATTERING TIMES%%%%%%%%%%%%%%%%%%%%%%%%%%%%%%%%%%%%%%%%%%%%%%%%%%%%%%%%%%%%%%%5
\section{Scattering times}
\label{Sec:scattering}

Depending on the temperature and the concentrations of the dopants, resistivity in a WSM
can be dominated by one of two quasiparticle scattering mechanisms.
For sufficiently large amount of disorder, resistivity is expected to come from electron scattering on screened
impurities. However, quasiparticles also can decay due to electron-electron interactions, which
leads to a finite resistivity even in a disorder-free sample\cite{AbrikosovBeneslavskii,BurkovHookBalents}.

{\it Elastic scattering} on screened impurities is characterised by the transport scattering time, in
the Born approximation given by
\begin{equation}
	\tau_{\text{tr}}^{-1}(\bp)=\pi n_{\text{imp}}\rho\left(v|\bp|\right)
	\int \frac{do}{4\pi}(1-\cos^2\theta)\left|u[2p\sin({\theta}/{2})]\right|^2,
	\label{TranspGen}
\end{equation}
where $u(\bk)=u(|\bk|)$ is the Fourier-transform of the screened impurity potential, Eq.~(\ref{Screened}),
and $\rho(\varepsilon)=\varepsilon^2/(2\pi v^3)$ is the density of electron states at energy $\varepsilon$
(per spin and per Weyl point).
For the screened Coulomb potential we find, using Eqs.~(\ref{TranspGen}) and (\ref{Screened}),
\begin{equation}
	\tau_{\text{tr}}^{-1}(\bp)=\frac{2\pi \alpha^2 n_{\text{imp}}v}{p^2}
	\left[
	\left(1+\frac{1}{2\lambda^2p^2}\right)
	\ln\left(1+4p^2\lambda^2\right)-2
	\right],
	\label{TauTr}
\end{equation}
with the screening radius $\lambda(\mu,T)$ given by Eq.~(\ref{lambda}).

The characteristic momentum of the particles that contribute to the conductivity
and the screening radius can be estimated as $p\sim \max(\mu,T)/v$ and
$\lambda\sim (g\alpha)^{-\frac{1}{2}}v/\max(\mu,T)$ [Eq.~(\ref{lambda})], respectively.
Because of the large values of the parameter $p\lambda\sim(g\alpha)^{-\frac{1}{2}}\gg1$,
we neglect in what follows the dependence of the argument of the logarithm in Eq.~(\ref{TauTr})
on the momentum $p$, $\ln\left(1+4p^2\lambda^2\right)\rightarrow\frac{1}{2}|\ln(g\alpha)|$.

{\it Inelastic scattering.}
In the limit of an undoped material ($\mu\rightarrow0$)
the rate of scattering due to electron-electron interactions\cite{AbrikosovBeneslavskii,CompleteRubbish}
for a quasiparticle with momentum $\bp$
is given by (up to logarithmic prefactors)
\begin{equation}
	\tau^{-1}_{ee}(\bp)\sim g\alpha^2 v|\bp|.
	\label{TauEE}
\end{equation}
and decreases for a finite chemical potential.

{\it Disorder- vs. interactions-dominated transport.}
Eqs.~\eqref{TauTr} and \eqref{TauEE} show that the rate of inelastic scattering exceeds that of
scattering on screened impurities only for quasiparticles with sufficiently high energies
\begin{equation}
	\varepsilon\gtrsim T_{\text{imp}}\equiv v(n_{\text{imp}}/g)^\frac{1}{3},
\end{equation}
where we have omitted the logarithmic factor of Eq.~(\ref{TauTr}).
Because the chemical potential satisfies the condition
$\mu\lesssim v(n_{\text{imp}}/g)^\frac{1}{3}$, as follows from Eqs.~\eqref{neutrality} and \eqref{nofmu},
resistivity is dominated by the scattering due to electron-electron interactions only
at sufficiently high temperatures $T\gg T_{\text{imp}}$, where $T=T_{\text{imp}}$ thus determines
the boundary between the ``interactions-dominated'' and ``thermal screening'' regimes
in Fig.~\ref{Fig:Regimes}.

%%%%%%%%%%%%%%%%%%%%%%%%%%%%%%%%%%%%%%%%%%%%%%
\subsection*{Effective single-particle model.}

In the rest of the paper we evaluate the conductivity $\sigma(T,\omega,n_A,n_D)$
of a doped WSM, focussing on the disorder-dominated transport with small fluctuations
of the screened impurity potential, i.e. in the
``thermal screening'' and ``metallic'' regions in the diagram
in Fig.~\ref{Fig:Regimes}. Also, we analyse the crossover of conduction to the other regimes
on the boundaries of these regions.

The problem
then can be considered effectively single-particle and described by the Hamiltonian
\begin{align}
	\cH^{\text{eff}} &=\cH_0+\cH_{\text{imp}}^{\text{eff}} \\
	\cH_{\text{imp}}^{\text{eff}}
	&=\sum_i\int \hpsi^\dagger(\br)\frac{e^2}{\kappa|\br-\br_i|}e^{-\frac{|\br-\br_i|}{\lambda}}\hpsi(\br)\: d^3\br,
\end{align}
where $\cH_0$ is the Hamiltonian of free Weyl fermions, and $\cH_{\text{imp}}$ describes
the effective disorder potential, with the screening
length $\lambda$ given by Eq.~(\ref{lambda}).

\section{Conductivity}

\subsection{General expressions}
\label{Sec:GenExpr}

The conductivity of a disordered WSM is given by the Kubo-Greenwood formula
\begin{eqnarray}
	\sigma(\omega)=
	(2\pi\omega)^{-1}v^2g\int d\varepsilon\left[f_0(\varepsilon)-f_0(\varepsilon+\omega)\right]
	\nonumber\\
	\int d^3\br^\prime\:\:\mathrm{Tr}\left<\hat\sigma_x \hat G^A(\varepsilon+\omega,\br,\br^\prime)
	\hat\sigma_x \hat G^R(\varepsilon,\br^\prime,\br)\right>_{\text{dis}},
	\label{Kubo}
\end{eqnarray}
where $v\hsigma_x$ is the velocity operator along the $x$ axis; $\hat G^A(\varepsilon,\br,\br^\prime)$
and $\hat G^R(\varepsilon+\omega,\br,\br^\prime)$ are, respectively, the advanced and retarded Green's
functions, $2\times 2$ matrices in the pseudospin space, and $\langle\ldots\rangle$ denotes the averaging
with respect to disorder realisations.

In the limit of weak disorder,
\begin{equation}
	\tau_{\text{tr}}^{-1}[\max(\mu,T)/v]\ll\max(\mu,T),
	\label{WeakDisCondition}
\end{equation}
the disorder-averaged correlator in Eq.~\eqref{Kubo} can be conveniently
evaluated using a perturbative diagrammatic
technique\cite{AGD}.

%%%%%%%%%%%%%%%%%%%%%%%%%%%%%%%%%%%%%%%%%%%%%%%%%%%%%%%%%%%%%%%%%%%%%%%%%%%%%%%%%%%%%%%%%%%%%%%%%%%%%%%%%%%
%%%%%%%%%%%%%%%%%%%%%%%%%%%%%%FIGURE%%%%%%%%%%%%%%%%%%%%%%%%%%%%%%%%%%%%%%%%%%%%%%%%%%%%%%%%%%%%%%%%%%%%%%%
\begin{figure}[ht]
	\centering
	\includegraphics[width=0.4\textwidth]{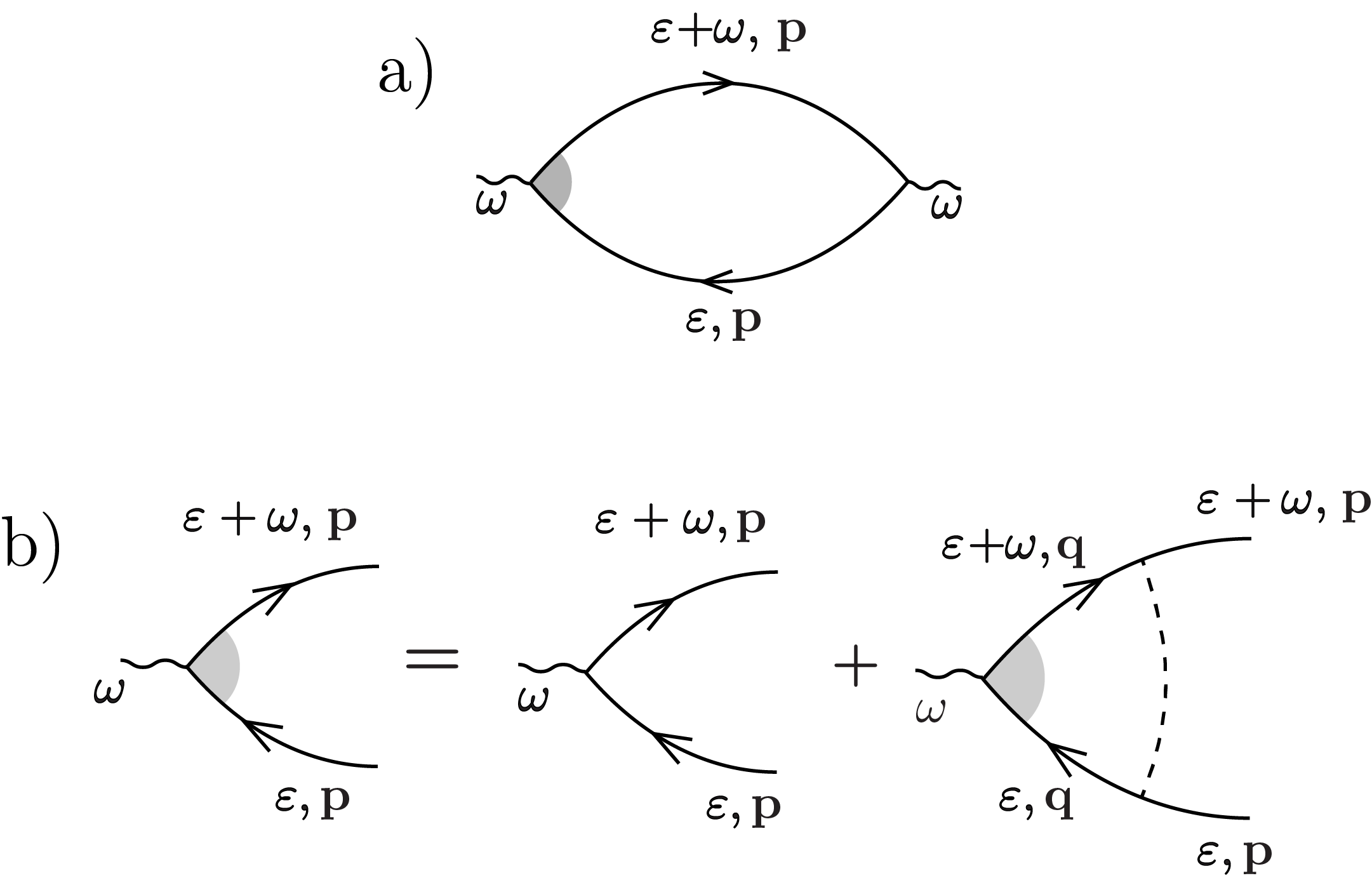}
	\caption{
	\label{Diagrams}
	Diagrams for the conductivity.
	a) Current-current correlator. b) Dyson equation for the current vertex renormalised by disorder.
	Solid line is a fully disorder-averaged propagator of Weyl fermions.
	}
\end{figure}
%%%%%%%%%%%%%%%%%%%%%%%%%%%%%%%%%%%%%%%%%%%%%%%%%%%%%%%%%%%%%%%%%%%%%%%%%%%%%%%%%%%%%%%%%%%%%%%%%%%%%%%%%%%%
%%%%%%%%%%%%%%%%%%%%%%%%%%%%%%%%%%%%%%%%%%%%%%%%%%%%%%%%%%%%%%%%%%%%%%%%%%%%%%%%%%%%%%%%%%%%%%%%%%%%%%%%%%%%

The conductivity is given by the Drude contribution (shown diagrammatically
in Fig.~\ref{Diagrams}),
which can be expressed in terms of the renormalised current vertex $\bJ(\bp,\omega,\varepsilon)$
and the Fourier-transforms $\left<G^A(\varepsilon+\omega,\bp)\right>_{\text{dis}}$
and $\left<G^R(\varepsilon,\bp)\right>_{\text{dis}}$ of the disorder-averaged Green's functions as
\begin{eqnarray}
	\sigma(\omega)=
	\frac{v g}{\omega}\int \frac{d\varepsilon}{2\pi}\:[f_0(\varepsilon)-f_0(\varepsilon+\omega)]
	\nonumber
	\\
	\int \frac{d^3\bp}{(2\pi)^3}\mathrm{Tr}
	\left[
	\hat J_x(\bp,\omega,\varepsilon)
	\left<G^A(\varepsilon+\omega,\bp)\right>_{\text{dis}}
	\right.
	\nonumber
	\\
	\left.
	\hat\sigma_x
	\left<G^R(\varepsilon,\bp)\right>_{\text{dis}}
	\right].
	\label{Drude}
\end{eqnarray}

In this paper we focus on the range of frequencies $\omega$, that
can be used to probe experimentally the quasiparticle scattering rate $\tau_{\text{tr}}^{-1}$
and are, therefore, smaller than the characteristic quasiparticle energy,
\begin{equation}
	\omega\ll\max(T,\mu).
	\label{SmallOmega}
\end{equation}
In the opposite limit, $\omega\gg\max(T,\mu)$, the WSM may be considered as a system of free
Dirac fermions; the electron dynamics is collisionless [cf. the condition~\eqref{WeakDisCondition}]
and the conductivity
$\sigma(\omega)=\frac{ge^2}{24v}\omega$ (see, e.g., Ref.~\onlinecite{CompleteRubbish}) is determined by
the interband electromagnetic-field-induced transitions between the valence and the conduction bands.

Under the condition \eqref{SmallOmega},
Eq.~\eqref{Drude} simplifies to (see Appendix~\ref{App:cond} for details)
\begin{equation}
	\sigma(\omega,\mu,T)=-\frac{e^2v^2g}{3\omega}\int\limits_{-\infty}^\infty\frac{\ve^2d\ve}{2\pi}
	\frac{f_0(\varepsilon+\omega)-f_0(\varepsilon)}{\tau_{\text{tr}}^{-1}(\varepsilon)-i\omega}.
\label{cond-final}
\end{equation}
From Eq.~(\ref{cond-final}) we find, after some algebra
(see Appendix~\ref{Sec:derivingF}),
\begin{equation}
   \sigma=-\frac{g}{i\omega}
   \frac{e^2T^2}{24\pi^2v}
   F\left[\frac{\alpha}{T}\left(\frac{2\pi n_{\text{imp}} v^3}{\omega}\right)^\frac{1}{2}
   \left|\ln{(g\alpha)}\right|^\frac{1}{2},\frac{\mu}{T},\frac{\omega}{T}\right],
   \label{MainAnswer}
\end{equation}
where
%\begin{align}
%  &F[\beta,\gamma]=
% \frac{4\pi^2}{3}-4(i\beta^2-\gamma^2)+\frac{\beta^3}{\pi}\frac{1-i}{\sqrt{2}}
%  \nonumber\\
%  &\times
%   \Big[\psi^\prime\Big(\frac{1}{2}-\frac{\beta}{2\pi\sqrt{2}}-\frac{i\beta}{2\pi\sqrt{2}}-\frac{i\gamma}{2\pi}\Big)
%   \nonumber\\
%   &-\psi^\prime\Big(\frac{1}{2}+\frac{\beta}{2\pi\sqrt{2}}+\frac{i\beta}{2\pi\sqrt{2}}-\frac{i\gamma}{2\pi}\Big)\Big]
%   \nonumber\\
%   &-\frac{\pi \beta^3}{\sqrt{2}}(1-i)\frac{1}{\cos^2\frac{1}{2}\left(\beta\frac{1+i}{\sqrt{2}}+i\gamma\right)},
%   \label{Fdef}
%\end{align}

\begin{align}
  &F(\beta,\gamma,\Omega)=\frac{4}{3}(\pi^2+\Omega^2)-4(i\beta^2-\gamma^2)-4\gamma\Omega+\frac{2\sqrt{i}\beta^3}{\Omega}\nonumber\\
  &\times\Bigg[\psi\left(\frac{1}{2}-\frac{i\gamma+\sqrt{i}\beta}{2\pi}\right)-
  \psi\left(\frac{1}{2}-\frac{i[\gamma-\Omega]+\sqrt{i}\beta}{2\pi}\right)\nonumber\\
   &+\psi\left(\frac{1}{2}-\frac{i[\gamma-\Omega]-\sqrt{i}\beta}{2\pi}\right)-
   \psi\left(\frac{1}{2}-\frac{i\gamma-\sqrt{i}\beta}{2\pi}\right)\nonumber\\
   &+2\pi i\bigg(\frac{1}{e^{i\sqrt{i}\beta-\gamma}+1}-\frac{1}{e^{i\sqrt{i}\beta-\gamma+\Omega}+1}\bigg)\Bigg]
 \label{Fdef}
\end{align}
with $\psi(x)$ being the digamma function\cite{AbramowitzStegun}, and the
ratio $\mu/T$ of the chemical potential to the temperature given by
[as follows from Eqs.~\eqref{neutrality} and \eqref{nofmu}]
\begin{equation}
	\frac{\mu}{T}=
	\frac{2\pi}{\sqrt{3}}\sinh
	\left[
	\frac{1}{3}\arcsinh
	\left(\frac{9\sqrt{3}}{\pi}\frac{v^3}{g}\frac{n_D-n_A}{T^3}\right)
	\right].
	\label{MuTRatio}
\end{equation}

%chemical potential $\mu$ related to the concentrations $n_D$ and $n_A$ of donors
%and acceptors as
%\begin{gather}
%    n_D-n_A=\frac{g}{6\pi^2v^3}(\mu^3+\pi^2\mu T^2).
%    \label{nDnAmu}
%\end{gather}

Eqs.~(\ref{MainAnswer}), (\ref{Fdef}), and (\ref{MuTRatio})
is our main result for the conductivity $\sigma(T,\omega,n_A,N_D)$
of a WSM as a function of temperature $T$, frequency $\omega$,
and the impurity concentrations $n_A$ and $n_D$.
In what immediately follows we analyse
the limiting regimes of low temperatures, frequencies, and the amount of disorder and
discuss the crossover behaviour between these regimes.

%%%%%%%%%%%%%%%%%%%%%%%%%%%%%%%%%%%%%%%%%%%%%%%%%%%%%%%%%%%%%%%%%%%%%%%%%%%%%%%%%%%%%%%%%%%%%%%%%%%%%%%5

\subsection{Metallic regime ($T\ll\mu$)}
\label{Subsec:LowT}

For low temperatures, $T\ll\mu\sim v\left|\frac{n_D-n_A}{g}\right|^\frac{1}{3}$,
the transport properties of the WSM are similar to those of a usual metal\cite{Abrikosov:metals};
conduction comes from low-energy excitations near the Fermi energy
$\varepsilon_F=\left[6\pi^2v^3(n_D-n_A)/g\right]^\frac{1}{3}\gg T$.

From Eqs.~(\ref{MainAnswer}) and (\ref{Fdef}) we find the conductivity in this regime
\begin{equation}
	\sigma\approx \left(\frac{g}{6\pi^2}\right)^{1/3}e^2v\frac{|n_D-n_A|^{2/3}}{\tau^{-1}_{tr}-i\omega}
	\label{SigmaLowT}
\end{equation}
with the transport scattering time given by
\begin{equation}
	\tau^{-1}_{tr}=2\pi\left(\frac{g}{6\pi^2}\right)^{2/3} \frac{n_{\text{imp}} v}{|n_D-n_A|^{2/3}}\alpha^2|\ln({g\alpha})|.
	\label{TauMetal}
\end{equation}

The conductivity is weakly temperature-dependent and has the frequency dependency
of a usual metal\cite{Abrikosov:metals},
$\sigma\propto(\tau^{-1}_{tr}-i\omega)^{-1}$.

%%%%%%%%%%%%%%%%%%%%%%%%%%%%%5
\subsection{Regime of thermal screening, $\mu\ll T$ }

In the temperature interval
\begin{equation}
	(|n_D-n_A|/g)^\frac{1}{3}v\ll T\ll (n_{\text{imp}}/g)^\frac{1}{3}v,
\end{equation}
the chemical potential $\mu$ is small compared to $T$, and
transport in the WSM
is determined by electrons thermally excited from the vallence to conduction band.

From Eqs.~(\ref{MainAnswer}) and (\ref{Fdef}) we find in this regime
\begin{align}
   \sigma(\omega,T)=\!\frac{ge^2  T^2}{24\pi^2n_{\text{imp}}v^4}\!
   \begin{cases}
      \!\frac{14\pi^3T^2}{15\alpha^2 |\ln({g\alpha})|},\!\!
      &\omega\ll\!\frac{\alpha^2 |\ln (g\alpha)| n_{\text{imp}}v^3}{T^2},\\
      \!\frac{4n_{\text{imp}}v^3\pi^2i}{3\omega},\!\!&\omega\gg\!\frac{\alpha^2|\ln (g\alpha)|n_{\text{imp}}v^3}{T^2}.
   \end{cases}
   \label{LowDopingCases}
\end{align}

The limiting cases of low and high frequencies, Eq.~\eqref{LowDopingCases}, can be summarised by
the interpolation formula
\begin{equation}
   \sigma(\omega,T)=\frac{ge^2  T^2}{18v}\frac{1}{\tau_0^{-1}-i\omega}
   \label{SigmaLowDop}
\end{equation}
with the effective scattering time
\begin{equation}
	\tau_0^{-1}=\frac{10}{7}\frac{\alpha^2n_{\text{imp}}v^3}{\pi T^2}|\ln (g\alpha)|.
	\label{Tau0}
\end{equation}

Eqs.~(\ref{SigmaLowDop}) and (\ref{Tau0}) resemble
the usual-metal result\cite{Abrikosov:metals} $\sigma(\omega)\propto[\tau^{-1}_{tr}-i\omega]^{-1}$
with an effective transport scattering time $\tau_{\text{tr}}=\tau_0$, Eq.~\eqref{Tau0}.
Indeed, Eqs.~(\ref{SigmaLowDop}) and (\ref{Tau0}) can be understood as a result of the averaging
of a metallic conductivity over an interval of energies $\varepsilon\sim T$ [cf. Eq.~\eqref{cond-final}].
Because the density of states and the transport scattering time [Eq.~\eqref{TauTr}] in a WSM
are strongly energy-dependent, the effective scattering time and the prefactor in
the averaged conductivity, Eq.~(\ref{SigmaLowDop}),
strongly depend on temperature.

For zero frequency, $\omega=0$, we reproduce the temperature
dependence $\sigma\propto T^4$, obtained in Ref.~\onlinecite{HwangSarma} for a weakly-doped WSM (for sufficiently
high temperatures).

At high frequencies, $T\gg\omega\gg\tau_0^{-1}$, the quasiparticle dynamics is collisionless,
hence the conductivity is imaginary and decreases with frequency as
$\sigma(\omega)\propto i/\omega$. We note, that this collisionless regime should be
contrasted with the collisionless regime at $\omega\gg T$
with a real conductivity\cite{CompleteRubbish} $\sigma(\omega)=\frac{ge^2}{24v}\omega$
that comes from the interband transitions.

{\it Crossover to the interactions-dominated regime.} For $T\sim T_{\text{imp}}\equiv[(n_A+n_D)v^3/g]^{1/3}$
the dc conductivity
\begin{equation}
	\sigma\sim \frac{T_{\text{imp}}}{\alpha}
\end{equation}
matches the conductivity\cite{CompleteRubbish,BurkovHookBalents}
$\sigma\sim T/\alpha$
of a disorder-free sample, because at $T\sim T_{\text{imp}}$ transport in a WSM crosses
over from the disorder-dominated to the interactions-dominated regime.

%%%%%%%%%%%%%%%%%%%%%%%%%%%%%%%%%
\subsection{dc Limit, $\omega\rightarrow0$.}
\label{Subsec:dcLim}

For very low frequencies,
\begin{equation}
   \omega\ll \alpha^2\frac{n_{\text{imp}}v^3 }{T^2}|\ln (g\alpha)|,
\end{equation}
Eqs.~(\ref{neutrality}) and (\ref{nofmu}) give

\begin{gather}
  \begin{split}
   \sigma(\omega,T)&\approx\frac{ge^2T^4}{48\pi^2n_{\text{imp}}v^3 \alpha^2|\ln (g\alpha)|}\Bigg[P_4\left(\frac{\mu}{T}\right)\\
   &+\frac{i\omega T^2}{2\pi n_{\text{imp}}v^3\alpha^2|\ln (g\alpha)|}P_6\left(\frac{\mu}{T}\right)\\
   &-\frac{\omega^2 T^4}{2(2\pi n_{\text{imp}}\alpha^2|\ln g\alpha|)^2}P_8\left(\frac{\mu}{T}\right)\Bigg]
   \end{split}
   \label{SigmaDC}
\end{gather}
where the ratio $\mu/T$ is given by Eq.~\eqref{MuTRatio}, and
\begin{gather}
 \begin{split}
  P_4(x)&=4x^4+8\pi^2 x^2+\frac{28\pi^4}{15},\\
    P_6(x)&=4x^6+20\pi^2 x^4+28\pi^4x^2+\frac{121\pi^6}{21},\\
     P_8(x)&=4\left(x^8+\frac{28\pi^2}{3} x^6+\frac{98\pi^4}{3}x^4+\frac{124\pi^6}{3}x^2+\frac{127\pi^8}{15}\right),
 \end{split}
\end{gather}

As necessary, in the limits $\mu/T\ll 1$ and $\mu/T\gg 1$ Eq.~\eqref{SigmaDC} reproduces respectively
Eq.~\eqref{SigmaLowDop} and \eqref{SigmaLowT} for $\omega=0$.

{\it Crossover to the minimal conductivity.}
For $T=0$, $|n_A-n_D|\sim g^\frac{1}{2}\alpha^\frac{3}{2} n_{\text{imp}}$
and for $T\sim g^\frac{1}{6}\alpha^\frac{1}{2}v(n_{\text{imp}}/g)^\frac{1}{3}$, $n_A=n_D$,
at the boundary of the ``minimal conductivity'' regime in Fig.~\ref{Fig:Regimes}, we find the value of the
conductivity
\begin{equation}
   \sigma_{\text{min}}\sim e^2 (gn_{\text{imp}})^\frac{1}{3},
   \label{MinCond}
\end{equation}
which, up to a prefactor of $g^\frac{1}{3}$ reproduces the minimal conductivity of a WSM obtained
in Ref.~\onlinecite{Skinner:WeylImp}. The conductivity is of the same order of magnitude, Eq.~(\ref{MinCond}),
in the entire ``minimal conductivity'' phase in Fig.~\ref{Fig:Regimes}, because
the depth
of electron and hole puddles in this regime significantly exceeds the temperature and the average chemical potential.

%%%%%%%%%%%%%%%%%%%%%%%%%%%%%%%%%%%%%%%%%%%%%%%%%%%%%%%%%%%%%%%%%%%%%%%%%%%%%%%%%%%%%%%%%

\section{Discussion}
\label{Sec:conclusion}

%{\it Summary.}
We have studied conductivity of a Weyl semimetal with donor and acceptor impurities.
Depending on the temperature $T$, and the concentrations $n_A$ and $n_D$ of acceptors
and donors, the material may be in one of four regimes of conduction, that are summarised
in Fig.~\ref{Fig:Regimes}.
We have evaluated the conductivity $\sigma(T,\omega,n_A,n_D)$ explicitly in the ``thermally screened''
and ``metallic'' regimes and discussed the crossover behaviour to the other regimes.

Although experimental realisations of 3D Dirac materials are rather few and recent,
it is possible to estimate the characteristic scales of temperatures,
doping levels, and frequencies for various regimes of transport in Fig.~\ref{Fig:Regimes},
using the parameters of the quasiparticle spectrum\cite{Liu:Cd3As2,Hasan:Cd3As2,Cava:Cd3As2,Yazdani:Cd3As2}
and the carrier density\cite{Yazdani:Cd3As2} in, e.g., the Dirac semimetal $\text{Cd}_3\text{As}_2$.

This material has the dielectric constant\cite{Cd3As2dielectric} $\kappa\approx 36$,
which for the Fermi velocity $v\sim 10^6\ \text{m}\cdot\text{s}^{-1}$ gives the effective
fine structure constant $\alpha\sim0.05$. Assuming the semimetal is not highly compensated,
the concentrations of the charge carriers and Coulomb impurities are of the same order of magnitude,
$|n_D-n_A|\sim n_{imp}\sim n$.
For the charge carrier density $n\sim10^{18}\,\text{cm}^{-3}$,
reported in Ref.~\onlinecite{Yazdani:Cd3As2}, both the chemical potential $\mu\approx 2000$K
and the characteristic temperature $T_{\text{imp}}\sim n^{1/3}v\sim 10^3$K of the interactions-dominated
scattering (see Fig.~\ref{Fig:Regimes}) significantly exceed the range of temperatures used in experiments.
In terms of transport properties, such material is thus rather similar to a metal
with the elastic scattering rate
[Eq.~\eqref{TauMetal}] $\tau_{\text{tr}}^{-1}\sim 10^{11}\ \text{Hz}\ (\sim 1\ \text{K})$.
In order to observe significant frequency dependency of the conductivity, the frequency has
to lie in the same range $\omega\sim 10^{11}\ \text{Hz}$ or higher. However, a weak frequency-dependent
correction to the dc conductivity [see Eq.~(\ref{SigmaDC})] can be also observed at smaller frequencies.

In order to drive the material away from the metallic regime,
the chemical potential has to be significantly reduced, which can by achieved by counterdoping,
i.e. by introducing extra impurities in order to achieve nearly equal concentrations of
donors and acceptors. For instance, achieving chemical potentials smaller than
the room temperature requires the level of compensation $|n_D-n_A|/n_{\text{imp}}\lesssim 10^{-3}$.
The conductivity in this ``thermal screened'' regime is strongly temperature-dependent,
$\sigma\propto T^2/(CT^{-2}-i\omega)$ with the effective elastic scattering rate [Eq.~\eqref{Tau0}]
at room temperature $\tau_0^{-1}=CT^{-2}\sim 10^{12}\ \text{Hz}$.

\section{Acknowledgements}

We are indebted to V.S.~Khrapai for useful discussions and remarks on the manuscript.
The work of SVS has been financially supported by the Alexander von Humboldt
Foundation through the Feodor Lynen Research Fellowship and by the NSF grants
DMR-1001240,
DMR-1205303, PHY-1211914, and PHY-1125844.
YaIR acknowledges financial support from the Ministry of
Education and Science of the Russian Federation under the
``Increase Competitiveness'' Programme of NUST "MISIS" (No. K2-2014-015),
from the Russian Foundation for Basic Research under the
project~14-02-00276, and from the Russian Science Support Foundation.

%%%%%%%%%%%%%%%%%%%%%%%%%%%%%APPENDIX%%%%%%%%%%%%%%%%%%%%%%%%%%%%%%%

\appendix

\section{Renormalised current vertex and conductivity}
\label{App:cond}

The renormalised current vertex, $\bJ(\omega,\bp,\varepsilon)$, satisfies the Dyson equation,
shown diagrammatically in Fig.~\ref{Diagrams}b,
\begin{align}
	\hat\bJ(\bp,\omega,\varepsilon)=ev\hbsigma
	\nonumber\\
	+ \int\frac{d^3\bq}{(2\pi)^3}|u(\bp-\bq)|^2
	\hat G^A(\varepsilon+\omega,\bq)\hat\bJ(\bq,\omega,\varepsilon)\hat G^R(\varepsilon,\bq),
	\label{DysonJ}
\end{align}
where $\bp$ is the fermionic momentum, that goes in and out of the vertex $\bJ(\bp,\omega)$
(Fig.~\ref{Diagrams}b), $\varepsilon$ is the incoming fermionic energy,
and $\omega$ is the frequency of the photon incoming in the vertex;
$u(\bk)$ is the Fourier-transform of the screened impurity potential; $ev\hbsigma$ is the
bare (non-renormalised) current vertex;
\begin{equation}
	\hat G^{R,A}(\varepsilon,\bp)=\frac{\varepsilon+v\hbsigma\bp}
	{\left[\varepsilon\pm\frac{i}{2\tau(\bp)}\right]^2-\left[vp\mp\frac{i}{2\tau_1(\bp)}\right]^2}
	\label{GRAexplicit}
\end{equation}
are the disorder-averaged advanced and retarded Green's functions where we have introduced the characteristic
scattering rates
\begin{align}
	\tau^{-1}(\bp)=\pi n_{\text{imp}}\rho\left(v|\bp|\right)
	\int \frac{do}{4\pi}\left|u[2p\sin({\theta}/{2})]\right|^2,\\
	\tau_1^{-1}(\bp)=\pi n_{\text{imp}}\rho\left(v|\bp|\right)
	\int \frac{do}{4\pi}\left|u[2p\sin({\theta}/{2})]\right|^2\cos\theta,
\end{align}
where $\int do\ldots$ is the integration with respect to solid angle.

In the limit of weak disorder and small frequencies under consideration,
only momenta $\bp$ close\cite{AGD} to the mass shell $p=\varepsilon/v$ contribute
to the response of quasiparticles with energy $\varepsilon$, encoded
by the second line of Eq.~\eqref{Kubo}. Accordingly, it is sufficient to consider the current
vertex in Eq.~\eqref{DysonJ} only with momenta $\bp$ close to $\varepsilon/v$.
Because of the smallness of the frequency $\omega$ the integration with
respect to momenta $\bq$ in Eq.~(\ref{DysonJ}) also can be assumed confined to a narrow
shell of momenta near the surface $q=\varepsilon/v$.

Under these assumptions
we look for the solution of the Dyson equation (\ref{DysonJ}) in the form
\begin{align}
	\hat\bJ(\bp,\omega,\varepsilon)=
	J_1(\omega)\: \hbsigma
	+J_2(\omega)\: v\bn
	+J_3(\omega)\: v^2{\bn(\hbsigma\cdot\bn)}
	\label{CurrentAnsatz}
\end{align}
with $J_1,\ldots,J_3$ being scalar functions and $\bn=\bp/p$.

Plugging the ansatz (\ref{CurrentAnsatz}) into the Dyson equation (\ref{DysonJ}) for the current vertex
and performing the momentum integration with $\omega\ll\varepsilon$
yields
\begin{align}
    J_1=ev+\frac{J_1+J_2+J_3}{2}\frac{\tau^{-1}_{\text{tr}}}{\tau^{-1}+\tau^{-1}_1-i\omega},
    \label{J1}
    \\
   J_2=\frac{(J_1+J_2+J_3)\tau^{-1}_1}{\tau^{-1}+\tau^{-1}_1-i\omega},
   \\
   J_3=\frac{(J_1+J_2+J_3)\left(\frac{1}{2}\tau^{-1}-\frac{3}{2}\tau^{-1}_{\text{tr}}\right)}
   {\tau^{-1}+{\tau_1}^{-1}-i\omega}.
   \label{J3}
\end{align}
From Eqs.~\eqref{J1}-\eqref{J3} we find the sum
\begin{gather}
   \label{vertex-sol}
 	J_1+J_2+J_3=
   ev\frac{\tau^{-1}+\tau^{-1}_1-i\omega}{\tau^{-1}_{\text{tr}}-i\omega},
\end{gather}
in terms of which the conductivity \eqref{Drude} in the limit $\omega\ll\varepsilon$
can be rewritten, using Eq.~(\ref{GRAexplicit}), as
\begin{gather}
    \begin{split}
  \sigma(\omega)=-\frac{ge^2}{3\pi v}\int\limits_{-\infty}^\infty\frac{\ve^2d\ve}{2\pi}
  %\partial_\ve f_0(\ve)
  \frac{f_0(\varepsilon)-f_0(\varepsilon+\omega)}{\omega}
  \\
  \frac{J_1(\omega)+J_2(\omega)+J_3(\omega)}{\tau^{-1}(\ve)+\tau^{-1}_1(\ve)-i\omega}.
  \end{split}
  \label{CondAlmost}
\end{gather}
Using Eqs.~(\ref{vertex-sol}) and (\ref{CondAlmost}) we arrive at Eq.~(\ref{cond-final}).

\section{General expression for conductivity}
\label{Sec:derivingF}

In this Section we present a detailed derivation of expressions (\ref{MainAnswer})
and (\ref{Fdef}) from Eq.~(\ref{cond-final}).

By introducing the notations
\begin{align}
	\gamma &=\frac{\mu}{T},\ \ \Omega=\frac{\omega}{T}
	\\
	a^2 &=\frac{2\pi\alpha^2 n_{\text{imp}}v^3}{T^2\omega}
	\ln\left(1+4p^2\lambda^2\right)
	\\
	\mathcal{D}_{x,\Omega}f(x) & \equiv\frac{f(x)-f(x-\Omega)}{\Omega}
\end{align}
and neglecting the dependence of the logarithm on its argument,
$\ln\left(1+4p^2\lambda^2\right)\rightarrow\frac{1}{2}|\ln (g\alpha)|$,
Eq.~(\ref{cond-final}) is reduced to Eq.~(\ref{MainAnswer}) with
\begin{gather}
    \begin{split}
    F(\beta,\gamma,\Omega)&=4\int\limits_{-\infty}^\infty\Big(\frac{1}{e^{x-\gamma-\Omega}+1}-\frac{1}{e^{x-\gamma}+1}\Big)
    \frac{dx\: x^4}{x^2+ia^2}\\
    \equiv&-4\mathcal{D}_{\gamma,\Omega}\int\limits_{-\infty}^\infty\frac{x^4}{e^{x-\gamma}+1}\frac{dx }{x^2+ia^2}.
   \end{split}
   \label{Finit}
\end{gather}

To evaluate the integral (\ref{Finit}) we split it into two parts:
$F(a,\gamma)\equiv 4F_I(a,\gamma,\Omega)- 4a^4 F_{II}(a,\gamma,\Omega)$,
\begin{gather}
   F_I(a,\gamma,\Omega)=\frac{\pi^2+\Omega^2}{3}-\gamma^2+ia^2+\gamma\Omega,
   \label{FI}
\end{gather}
\begin{gather}
 \begin{split}
   &F_{II}(a,\gamma,\Omega)=-\mathcal{D}_{\gamma,\Omega}\int_{-\infty}^\infty \frac{dx}{x^2+ia^2}\frac{1}{e^{x-\gamma}+1}\\
   &=-\mathcal{D}_{\gamma,\Omega}\int\limits_{-\infty}^\infty f_\gamma(x)\,dx
=-2\pi i\mathcal{D}_{\gamma,\Omega}\sum\limits_{n=0}^\infty\underset{x=x_n}{\text{res}}f_\gamma(x)\\
&-2\pi i\mathcal{D}_{\gamma,\Omega}\underset{x=i\sqrt{i}a}{\text{res}}f_\gamma(x),
  \end{split}
  \label{FII}
\end{gather}
where $\sqrt{i}=\frac{1+i}{\sqrt{2}}$.

The residue of the function $f(x)$ at $x=x_n=\gamma+i\pi(2n+1)$, $n=0,1,2,\ldots$ is given by
\begin{gather}
  \begin{split}
   &\underset{x=x_n}{\text{res}}f_\gamma(x)=
   -\frac{1}{x_n^2+ia^2}\\
   &=\frac{1}{2i\sqrt{i}a}\left[\frac{1}{x_n+i\sqrt{i}a}-\frac{1}{x_n-i\sqrt{i}a}\right],
  \end{split}
  \label{Res1}
\end{gather}
the residue at $x=i\sqrt{i}a$--
\begin{gather}
   \underset{x=i\sqrt{i}a}{\text{res}}f_\gamma(x)=\frac{1}{2i\sqrt{i}a}\frac{1}{e^{i\sqrt{i}a-\gamma}+1}.
   \label{Res2}
\end{gather}

Using Eqs.~(\ref{FII}), (\ref{Res1}), and (\ref{Res2}), we obtain
\begin{align}
   &F_{II}(a,\gamma,\Omega)=-\mathcal{D}_{\gamma,\Omega}\Bigg(
   \frac{\sqrt{i}}{2a}\sum\limits_{n=0}^\infty
   \left[\frac{1}{n+\frac{1}{2}-\frac{\sqrt{i}a}{2\pi}-\frac{i\gamma}{2\pi}}
   \right.
   \nonumber
   \\
   &
   \left.
   -\frac{1}{n+\frac{1}{2}+\frac{\sqrt{i}a}{2\pi}-\frac{i\gamma}{2\pi}}\right]
   +\frac{\pi}{\sqrt{i}a}\frac{1}{e^{i\sqrt{i}a-\gamma}+1}\Bigg).
   \label{sum1}
\end{align}

Using Eqs.~(\ref{FI}) and (\ref{FII}) and the definition of the digamma function\cite{AbramowitzStegun},
\begin{gather}
   \psi(z)=\sum\limits_0^\infty\left(\frac{1}{n+1}-\frac{1}{z+a}\right)-C,
\end{gather}
with $C$ being the Euler constant,
we arrive at Eq.~\eqref{Fdef}.

\end{document}